\documentclass[11pt,a4paper]{article}

\usepackage[utf8]{inputenc}
\usepackage[T1]{fontenc}
\usepackage{lmodern}
\usepackage[margin=1in]{geometry}
\usepackage{graphicx}
\usepackage{booktabs}
\usepackage{array}
\usepackage{tabularx}
\usepackage{hyperref}
\usepackage{xcolor}
\usepackage[numbers,sort&compress]{natbib}
\usepackage{enumitem}
\usepackage{amsmath}
\usepackage[font=small,labelfont=bf]{caption}
\usepackage{tikz}
\usetikzlibrary{arrows.meta,calc,positioning}

\hypersetup{
  colorlinks=true,
  linkcolor=blue!60!black,
  citecolor=blue!60!black,
  urlcolor=blue!60!black
}

\definecolor{figblue}{RGB}{92,112,153}
\definecolor{figgreen}{RGB}{102,128,107}
\definecolor{figorange}{RGB}{162,126,88}

\newcolumntype{L}[1]{>{\raggedright\arraybackslash}p{#1}}
\newcolumntype{Y}{>{\raggedright\arraybackslash}X}

\tikzset{
  paperbox/.style={
    rounded corners=2pt,
    line width=0.45pt,
    draw=black!58,
    fill=white,
    align=center,
    inner sep=5pt
  },
  surveybox/.style={
    rounded corners=2pt,
    line width=0.65pt,
    draw=black!72,
    fill=black!2,
    align=center,
    inner sep=6pt
  },
  rolebox/.style={
    rounded corners=2pt,
    line width=0.4pt,
    draw=black!50,
    fill=black!1,
    align=center,
    inner sep=4.5pt
  },
  bluebox/.style={paperbox, draw=figblue!70!black, fill=figblue!8},
  greenbox/.style={paperbox, draw=figgreen!70!black, fill=figgreen!10},
  orangebox/.style={paperbox, draw=figorange!78!black, fill=figorange!11},
  figarrow/.style={
    -{Latex[length=2.2mm,width=1.5mm]},
    line width=0.65pt,
    draw=black!70
  },
  figdash/.style={
    -{Latex[length=2.0mm,width=1.35mm]},
    line width=0.45pt,
    dashed,
    draw=black!60
  }
}

\title{Operational Evidence Gaps for LLMs in Fraud Detection and Trust-and-Safety Workflows}
\author{Keyur Gabani}
\date{}

\begin{document}

\maketitle

\begin{abstract}
LLMs are now proposed for fraud detection, scam investigation, content moderation, and other trust-and-safety workflows. Much of the public literature still evaluates them as models, with less attention to their behavior as components in operational pipelines. This creates a practical evidence question: what would justify placing an LLM inside a live workflow with latency, cost, escalation, human-review, and adversarial-risk constraints?

We address this question through a fraud-first survey of deployment evidence. We code 49 operationally relevant sources on LLM use in fraud detection, investigation support, content moderation, and cross-cutting robustness (18 fraud, 14 moderation, 17 cross-cutting), supplemented by 15 contextual references that establish the survey boundaries. These sources include systems, benchmarks, frameworks, and deployment-relevant surveys, not 49 production deployments.

The main finding is an evidence imbalance. Fraud supplies the largest task-specific portion of the coded corpus. The moderation papers, however, include more explicit public evidence on latency, cost, governance, and fairness. Among the 18 fraud and investigation sources, none report clean per-decision latency, per-decision dollar cost, or calibration evidence; most report offline task performance, retrieval gains, or case-study accuracy instead.

The survey contributes a role-and-evidence organizing frame, FORTE, for locating LLMs as classifiers, retrieval interfaces, explanation generators, reviewer assistants, agents, feature extractors, or escalation components. It also contributes a minimum deployment-evidence checklist covering latency budget, cost per decision, decision threshold, explanation integrity, and adversarial pressure. The resulting agenda identifies studies needed to support deployment claims for LLM-based fraud and trust-and-safety work.
\end{abstract}

\section{Introduction}

Trust and safety is often a production-ML domain. Fraud scoring, scam interception, content moderation, and investigation support frequently operate under adversarial pressure, tight error budgets, and reviewer queues that may have to clear in real time. In this setting, an LLM must be evaluated as a pipeline component under latency, cost, escalation, and audit constraints.

Large language models are now being proposed and evaluated for precisely these settings. Recent work applies LLMs to fraud detection pipeline designs \cite{FinFRERAG, FLAG, LLMSelfEvolvingRL}, retrieval-augmented decision support \cite{RAGPhoneFraud, LLMAuthFraud}, content moderation \cite{Aegis, SLMMod}, scam intelligence collection \cite{CASE}, explanation generation \cite{EXPLICATE}, and analyst-facing investigation workflows \cite{CoInvestigatorAI, FAAFramework}. Adjacent research on prompt injection, retrieval poisoning, explanation integrity, and abstention makes clear that operational use brings operational risk \cite{AgentSecBench, PoisonedRAG, SCRK, SHAPLLM}. Capability and reliability are tightly coupled in this domain.

Existing surveys do not line up neatly with this operational question. The LLM safety and security surveys reviewed here focus on the safety, privacy, alignment, or security \textit{of} LLMs \cite{SafeguardingLLMs, SafetyAtScale, LLMSecPriv}; those are valuable but answer a different question. Fraud and abuse surveys are closer, but still leave a distinct opening. Deep-learning fraud surveys \cite{DLFraudSurvey} are broad method reviews. Recent LLM financial-fraud reviews \cite{LLMFinancialFraudReview} focus on financial-fraud methodologies, implementation challenges, and domain coverage. Online-fraud AI/NLP surveys \cite{OnlineFraudAISLR} cover text-based scam and fraud detection at much larger scale, while abuse-detection lifecycle work \cite{KathADL} explains when LLMs enter abuse pipelines. This paper instead asks what operational role the LLM plays, what evidence supports that role, and where fraud-specific deployment evidence remains missing.

This survey focuses on LLMs proposed for or evaluated in fraud detection and adjacent trust-and-safety workflows. The organizing principle is what role the LLM plays inside a working system. That makes it possible to compare classifiers, retrieval interfaces, explanation generators, and agentic investigation workflows while keeping practical constraints visible: latency, cost, review queues, abstention, robustness, and cross-domain generalization.

Treating LLMs as tools for safety operations changes the scope of the survey. Relevant work must cover where the model sits in a workflow and how its output is evaluated, not only whether the model itself is aligned or secure. A moderation system with adaptive strictness \cite{FlexGuard}, a fraud review assistant that mixes trusted structured evidence with untrusted free-form text \cite{FAAFramework}, and an LLM-generated explanation pipeline for high-stakes decisions \cite{LLMExplDiv} all raise deployment questions that alignment or jailbreak surveys usually leave to the side.

The paper is both an evidence audit and a taxonomy. FORTE provides the reader-facing frame; the evidence matrix records the operational fields behind it: application category, role/use, pattern, evidence tags, and notes. Four recurring placement patterns describe where papers put LLMs in pipeline designs. Fraud evidence remains detection-heavy and deployment-light, especially for latency, cost, and calibration. Section~6 examines what evidence would make each placement supportable in practice, and Section~7 proposes nine studies. Three areas are particularly underdeveloped in the searched corpus and window: concept drift, procedural justice and appeals, and head-to-head cross-domain transfer.

\section{Related Work and Background}

The literature relevant to this survey sits in three largely disconnected bodies of work: surveys on the safety, alignment, and security of large language models; surveys on machine learning for fraud detection; and a growing set of applied papers using LLMs inside trust-and-safety workflows.

\paragraph{Surveys on LLM safety and security.} Several broad surveys on LLM safety, alignment, and security are closest to the risk side of this paper, but their center of gravity is different. \textit{Safeguarding LLMs} \cite{SafeguardingLLMs} maps guardrails and safety mechanisms. \textit{Safety at Scale} \cite{SafetyAtScale} extends the lens to large models and agentic systems. The \textit{LLM Security and Privacy} survey \cite{LLMSecPriv} catalogs threats and defenses against the model itself, and related reviews \cite{GuardiansOffenders, SafetyReasoningModels} sit in the same model-centric register. The operational question here is narrower: how an LLM behaves as one component inside a trust-and-safety pipeline.

\paragraph{Fraud and abuse surveys.} The closest neighbouring fraud surveys now form three distinct comparison points. \textit{Year-over-Year Developments in Financial Fraud Detection via Deep Learning} \cite{DLFraudSurvey} reviews 57 deep-learning fraud studies from 2019--2024, but LLMs receive limited treatment and the organizing logic is method-family coverage rather than deployment-role synthesis. Kumar and Shaik's 2026 systematic review of LLMs for financial fraud detection \cite{LLMFinancialFraudReview} is closer on model family and domain, covering 33 LLM-related financial-fraud studies across detection performance, architectures, computing requirements, implementation challenges, and future directions. Papasavva et al.'s online-fraud SLR \cite{OnlineFraudAISLR} is broader and more systematic, screening 2,457 records and analyzing 223 AI/NLP papers on text-based online fraud. These surveys establish that fraud detection is already well reviewed as an AI/NLP or financial-fraud topic. The gap for this paper is narrower: a fraud-first synthesis of LLM operational roles, deployment patterns, and evidence gaps across fraud and adjacent trust-and-safety workflows.

\paragraph{Applied work at the intersection.} Applied papers now sit between these survey traditions. FinFRE-RAG \cite{FinFRERAG} and RAG-based phone fraud systems \cite{RAGPhoneFraud} show LLMs as evidence-grounded decision support. FLAG \cite{FLAG} and DGP \cite{DGP} use LLMs to enhance graph-based fraud pipelines. Co-Investigator AI \cite{CoInvestigatorAI}, CASE \cite{CASE}, and the FAA Framework \cite{FAAFramework} push the literature into agentic investigation. Aegis \cite{Aegis}, FlexGuard \cite{FlexGuard}, SLM-Mod \cite{SLMMod}, and Policy-as-Prompt \cite{PolicyAsPrompt} illustrate moderation system designs and evaluations. Agent Security Bench \cite{AgentSecBench}, PoisonedRAG \cite{PoisonedRAG}, and explanation-divergence work \cite{LLMExplDiv} expose robustness and integrity problems. A PRISMA-following survey of LLM moderation \cite{LLMModSurvey} covers that subfield in detail, and two agent-security surveys \cite{AgentSecInfoFusion, AgentSecACMCS} map the broader threat model.

\paragraph{A concurrent lifecycle perspective.} Kath et al.\ \cite{KathADL} organize LLM use in abuse detection along a lifecycle perspective that includes data work, detection, review, appeals, auditing, and governance. Their stages and the FORTE lenses overlap on workflow placement, but the taxonomies answer different questions. Kath et al.\ ask \emph{when} an LLM is acting in an abuse pipeline. This paper asks \emph{what role} it plays: classifier, retrieval interface, explanation generator, reviewer assistant, agent, feature extractor, or escalation component. The two views fit together: a lifecycle stage says where work happens, and an operational role says what the LLM is allowed to do there.

These papers are individually valuable but scattered across fraud, moderation, security, explainability, and agent-systems venues. In the surveyed neighbour literature, we did not find a shared organizing framework for fraud-centered LLM roles and deployment evidence. Table~\ref{tab:landscape} summarizes the distinction.

\begin{table}[!htbp]
\centering
\caption{Survey Landscape Comparison}
\label{tab:landscape}
{\setlength{\tabcolsep}{4pt}
\renewcommand{\arraystretch}{1.12}
\footnotesize
\begin{tabularx}{\textwidth}{L{3.15cm}Y Y Y}
\toprule
\textbf{Strand and examples} & \textbf{Organizing question} & \textbf{Covers well} & \textbf{Gap left for this survey} \\
\midrule
Safety / security \textit{of} LLMs \cite{SafeguardingLLMs, SafetyAtScale, LLMSecPriv} & How can models be made safe, private, aligned, or secure? & Guardrails, model risks, red-teaming & Operational roles, escalation, cost, and reviewer workflows \\
Fraud / online-fraud surveys \cite{DLFraudSurvey, LLMFinancialFraudReview, OnlineFraudAISLR} & How have AI or LLM methods for fraud evolved? & Datasets, model families, domain coverage & LLM roles, deployment evidence, and trust-and-safety comparison \\
LLM moderation \cite{LLMModSurvey, ModAccuracyLegitimacy} & How are LLMs used for moderation? & Architectures, legitimacy, policy framing & Fraud workflows, investigation, and cross-domain transfer \\
Agent security \cite{AgentSecInfoFusion, AgentSecACMCS} & What threats do LLM agents face? & Attack taxonomies and agent threat models & Trust-and-safety-specific deployment constraints \\
Applied T\&S papers \cite{FinFRERAG, CoInvestigatorAI, FLAG, SLMMod} & How are LLMs inserted into concrete tasks? & Systems, role diversity, case studies & Shared role taxonomy and evaluation standards \\
\textbf{This survey} & What roles do LLMs play in fraud-first T\&S workflows, and what evidence supports those roles? & FORTE roles, four placement patterns, evaluation-gap analysis & Large-scale bibliometric exhaustiveness and full production-deployment audit \\
\bottomrule
\end{tabularx}
}
\end{table}

The TELUS Digital report \cite{TELUS2025} confirms that industry adoption of AI for fraud and moderation is accelerating, but it serves here as motivation, not as scholarly backbone. Similarly, the LLMs for Explainable AI survey \cite{LLMsXAISurvey} and EXPLICATE \cite{EXPLICATE} help establish why explanation generation matters in fraud review, without resolving the synthesis question this paper addresses.

\paragraph{Entry points for new readers.} Readers new to this area can start with three papers that represent the survey's main evidence types: FAA Framework \cite{FAAFramework} for end-to-end agentic investigation, SLM-Mod \cite{SLMMod} for specialized smaller models under moderation constraints, and PoisonedRAG \cite{PoisonedRAG} or the indirect-prompt-injection competition \cite{IPIComp} for current adversarial risk. These are not the whole field, but they illustrate the paper's role, evidence, and trust axes.

\begin{figure}[t]
\centering
\begin{tikzpicture}[node distance=8mm and 8mm, every node/.style={font=\footnotesize}]
    \node[paperbox,text width=3.0cm,minimum height=1.65cm] (safety) {\textbf{Safety-of-LLMs surveys}\\[-0.5mm]\scriptsize Guardrails, alignment, and model security};
    \node[paperbox,text width=3.0cm,minimum height=1.65cm,right=of safety] (fraud) {\textbf{Fraud-ML surveys}\\[-0.5mm]\scriptsize Datasets, baselines, and method families};
    \node[paperbox,text width=3.0cm,minimum height=1.65cm,right=of fraud] (moderation) {\textbf{Moderation surveys}\\[-0.5mm]\scriptsize Architectures, legitimacy, and policy framing};
    \node[paperbox,text width=3.0cm,minimum height=1.65cm,right=of moderation] (applied) {\textbf{Applied deployment papers}\\[-0.5mm]\scriptsize Fraud, moderation, and agent workflows};

    \node[surveybox,text width=8.6cm,below=12mm of $(fraud.south)!0.5!(moderation.south)$] (gap) {\textbf{Operational gap}\\[-0.5mm]\scriptsize We did not find a survey that jointly organizes LLM roles, deployment patterns, and evaluation evidence for fraud and adjacent trust-and-safety work.};
    \node[greenbox,text width=4.0cm,below=10mm of gap] (thissurvey) {\textbf{This survey}\\[-0.5mm]\scriptsize FORTE with deployment and evaluation lenses};

    \coordinate (g1) at ($(gap.north west)!0.12!(gap.north east)$);
    \coordinate (g2) at ($(gap.north west)!0.37!(gap.north east)$);
    \coordinate (g3) at ($(gap.north west)!0.63!(gap.north east)$);
    \coordinate (g4) at ($(gap.north west)!0.88!(gap.north east)$);

    \draw[line width=0.5pt,draw=black!55] (safety.south) -- (g1);
    \draw[line width=0.5pt,draw=black!55] (fraud.south) -- (g2);
    \draw[line width=0.5pt,draw=black!55] (moderation.south) -- (g3);
    \draw[line width=0.5pt,draw=black!55] (applied.south) -- (g4);
    \draw[figarrow] (gap) -- (thissurvey);
\end{tikzpicture}
\caption{Survey landscape. Existing surveys cover safety \textit{of} LLMs, fraud-ML methods, or moderation independently. Applied papers are scattered across venues. This survey organizes the deployment-evidence gap.}
\label{fig:landscape}
\end{figure}
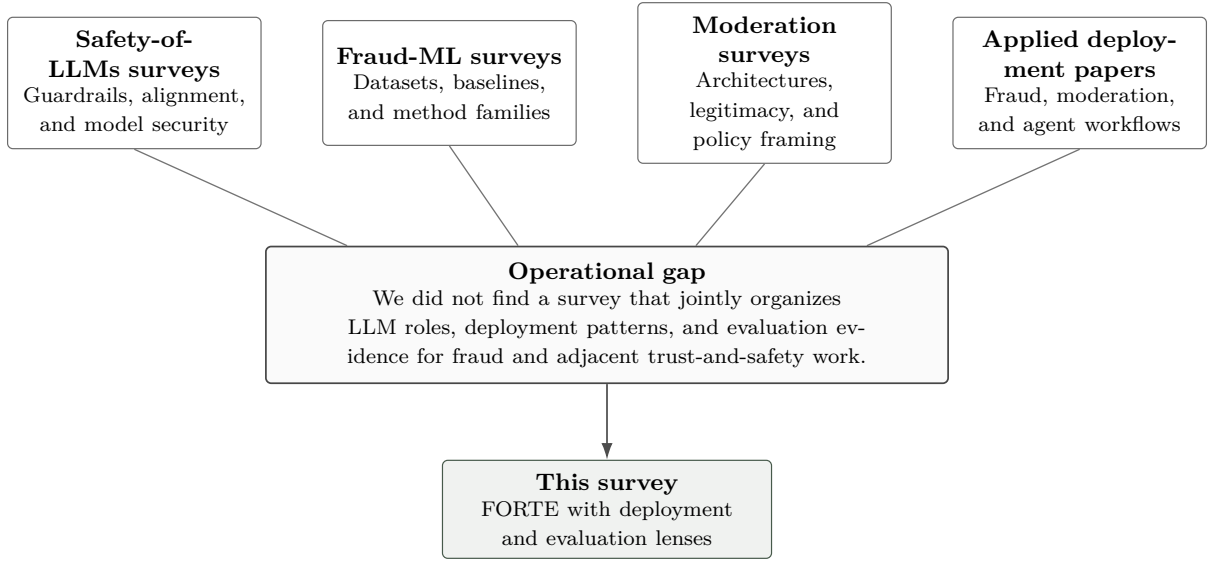

\section{Scope and Survey Methodology}

This paper is a structured narrative survey of LLMs proposed for or evaluated in fraud detection and adjacent trust-and-safety workflows. The methodological goal is a transparent, deployment-grounded synthesis, not a PRISMA-style systematic review. The scope is fraud-first: fraud detection (including scam and phishing) is the anchor domain because it is the largest task-specific part of the coded corpus; investigation and compliance are workflow stages; moderation and abuse-prevention papers serve as comparison areas, not equal-weight subfields.

Table~\ref{tab:protocol} summarizes the survey protocol.

\begin{table}[ht]
\centering
\caption{Survey Protocol}
\label{tab:protocol}
{\setlength{\tabcolsep}{4pt}
\renewcommand{\arraystretch}{1.05}
\footnotesize
\begin{tabularx}{\textwidth}{L{1.95cm}Y}
\toprule
\textbf{Element} & \textbf{Specification} \\
\midrule
Review type & Structured narrative survey with explicit collection and coding rules \\
Core question & How are LLMs used in fraud detection and adjacent T\&S workflows, and what evidence justifies those deployment roles? \\
Sources & arXiv, ACL Anthology, IEEE Xplore, ACM DL, Google Scholar, Semantic Scholar \\
Time window & 2023--2026 primary; earlier work for historical context only \\
Inclusion & Papers applying LLMs to T\&S tasks or deployment-style workflows, or offering surveys/benchmarks/methods directly relevant \\
Exclusion & Alignment-only work, generic jailbreak papers without T\&S relevance \\
Coding dimensions & Application area, LLM role, adaptation style, deployment pattern, evaluation style \\
Industry evidence & Admissible as deployment context only; does not anchor novelty claims \\
\bottomrule
\end{tabularx}
}
\end{table}

\paragraph{Corpus statistics.} The full bibliography contains 69 cited items. The coded May 2026 corpus contains 64 sources; five post-window sources from a July 2026 spot check are cited only as directional signals in Section~7 and are not counted in the May 2026 matrix. Of the 64 coded sources, 49 are operationally relevant sources that directly study, benchmark, or survey LLM use in trust-and-safety tasks: 18 in fraud, scam, phishing, and investigation; 14 in moderation and abuse; and 17 in cross-cutting robustness and trustworthiness. These categories are coded A, B, and C, respectively, in the evidence matrix. The 49-source set should not be read as 49 production deployments. The two core application categories contain 30 primary empirical, system, audit, or integration studies and 2 moderation survey/framework sources; the cross-cutting category contains 11 empirical, method, or benchmark papers and 6 surveys/frameworks. The remaining 15 coded sources are contextual or boundary-setting references: 5 safety-of-LLM surveys, 4 neighboring domain surveys, and 6 industry reports or incident writeups. The fraud and investigation category is the largest task-specific part of the corpus, which is why fraud anchors the survey; the moderation and abuse category grew most in the May 2026 update after a targeted scan of demographic-fairness audits and production moderation systems. Two placements required judgment: Kath et al.'s lifecycle survey \cite{KathADL} is counted as cross-cutting because it directly organizes LLM use inside abuse pipelines, and SHAPLLM \cite{SHAPLLM} is counted there because attribution-grounded explanation is a cross-cutting integrity method rather than a domain system. An ancillary evidence matrix included with the source bundle records the source-level coding used for these counts.

\begin{table}[ht]
\centering
\caption{Corpus and Evidence Coding Summary}
\label{tab:corpuscoding}
{\setlength{\tabcolsep}{4pt}
\renewcommand{\arraystretch}{1.12}
\footnotesize
\begin{tabularx}{\textwidth}{L{2.6cm}L{0.7cm}Y Y Y}
\toprule
\textbf{Category} & \textbf{N} & \textbf{Dominant roles} & \textbf{Operational evidence} & \textbf{Trust / robustness evidence} \\
\midrule
Fraud, scam, phishing, investigation & 18 & Classifier, RAG support, graph enrichment, explanation, investigation agent & 1 page-verified batch-runtime/deployment source; 0 clean per-decision latency sources; 1 token-use source; 3 human/user-facing sources; 0 dollar-cost sources & 1 drift-focused source; adversarial testing usually absent \\
Moderation and abuse & 14 & Adaptive moderator, policy prompt, small model, fairness audit, escalation predictor & 4 latency/cost or efficiency sources; 1 primary human-evidence source plus framework context & Fairness audits, monitoring, and policy-shift instrumentation \\
Cross-cutting trustworthiness & 17 & Prompt-injection benchmark, RAG poisoning, explanation integrity, abstention, agent security, lifecycle survey & Not primary & Dominant evidence type: robustness, integrity, abstention, agent security \\
Contextual / boundary sources & 15 & Safety-of-LLMs surveys, neighboring fraud surveys, industry incident reports & Context only & Context only \\
\bottomrule
\end{tabularx}
}
\end{table}

\paragraph{Corpus-construction process.} The corpus was built from the reference lists of the closest neighbour surveys (\cite{LLMModSurvey, DLFraudSurvey, AgentSecACMCS}) and expanded through Google Scholar and arXiv snowballing during April and May 2026. The six services in Table~\ref{tab:protocol} were used for discovery, retrieval, or bibliographic verification; the snowball searches themselves were run in Google Scholar and arXiv. Queries combined trust-and-safety vocabulary (\textit{fraud}, \textit{scam}, \textit{phishing}, \textit{moderation}, \textit{prompt injection}) with LLM vocabulary (\textit{large language model}, \textit{LLM agent}, \textit{RAG}, \textit{guardrail}). Candidate selection required judgment about what counted as \emph{LLM use}: roughly a dozen papers that used ``LLM'' to mean a BERT-style encoder were excluded, as were about a half-dozen papers whose only LLM use was offline data labeling without integration into a deployed pipeline.

The coding scheme emphasizes deployment role rather than model family. The role vocabulary is fixed to the seven values used throughout the paper: classifier, retrieval interface, explanation generator, reviewer assistant, agent, feature extractor, and escalation component. Finer free-text descriptors in the evidence matrix (adaptive moderator, graph enhancer, fairness corrector) are recorded as specializations of these seven roles. Patterns are coded as bounded automation, retrieval-grounded analyst support, hybrid escalation, or multi-step investigation support. Because this is a single-author survey, there is no inter-rater agreement statistic; coding consistency was handled by re-coding earlier papers after each new dimension was added.

\paragraph{Limitations.} This is not an exhaustive bibliometric review, and the corpus is shaped by where public literature is available. Four limitations matter most. First, traditional banking-side fraud, including card-not-present transaction scoring and account takeover at financial institutions, is weakly represented because most published LLM-fraud work is academic or telecom-side. Second, agentic-investigation evidence is dominated by three systems (Co-Investigator AI, CASE, FAA Framework), which is too narrow a foundation for broad generalization. Third, several moderation-deployment signals appear in industry posts and incident writeups rather than peer-reviewed publications. Fourth, the May 2026 update included a targeted scan for moderation-side fairness audits and production moderation systems, with no equally targeted scan for fraud-side latency/cost reporting; the corpus-construction queries covered fraud and scam vocabulary throughout and surfaced no fraud source reporting clean per-decision latency or dollar cost, but part of the measured fraud--moderation evidence imbalance could in principle reflect search effort rather than the literature. Because the evidence matrix was frozen after the May 2026 update, the numeric corpus counts should be read as a snapshot rather than as a continuously updated July 2026 bibliography. Public evidence about real deployment is thin because trust-and-safety work is difficult to publish without exposing details an adversary could use. These constraints shape the strength of the claims made below.

\section{FORTE: Organizing Frame for LLM Use in Fraud and Adjacent Trust-and-Safety}

The corpus spans direct fraud detection with FRAUDLLM \cite{FRAUDLLM}, adaptive moderation with Aegis \cite{Aegis}, multi-step investigation with the FAA Framework \cite{FAAFramework}, and attacks on retrieved evidence in PoisonedRAG \cite{PoisonedRAG}. Each places an LLM within, or evaluates a risk to, a trust-and-safety workflow where reliability has operational consequences.

FORTE is the reader-facing frame used in this paper to keep the coded dimensions visible. It names five parts of the survey:

\begin{itemize}[nosep]
\item \textbf{F}\,---\,Fraud-first anchoring of the corpus and the discussion (this section).
\item \textbf{O}\,---\,Operational deployment patterns where papers place LLMs inside pipeline designs (Section~5).
\item \textbf{R}\,---\,Roles the LLM plays at the placement: classifier, retrieval interface, explanation generator, reviewer assistant, agent, feature extractor, or escalation component.
\item \textbf{T}\,---\,Trust axes (robustness, explanation integrity, selective prediction) that cut across roles and patterns (Section~4.4).
\item \textbf{E}\,---\,Evaluation gap analysis between what is measured and what would matter in deployment (Section~6).
\end{itemize}

The evidence matrix implements this frame through practical fields:

\begin{enumerate}[nosep]
\item \textbf{Application category} (coded as \texttt{Bucket} in the matrix): whether a source belongs to the fraud anchor, the moderation comparison area, cross-cutting evidence, or contextual literature. This field supports the fraud-first scope represented by F but is not a one-to-one FORTE code.
\item \textbf{Pattern}: the operational placement used in the dominant-role count, corresponding to O.
\item \textbf{Role / use}: the LLM function, recorded as a specialization of the seven-role vocabulary, corresponding to R.
\item \textbf{Evidence tags and notes}: the trust and evaluation evidence used for the T and E parts of the analysis.
\end{enumerate}

Adaptation style and deployment setting are descriptors within these fields. The matrix remains a compact coding record, while the manuscript uses FORTE to connect deployment patterns in Section~5, trust axes in Section~4.4, and evidence gaps in Section~6.

\paragraph{How to read this frame.} FORTE gives fraud detection the most space because fraud is the largest task-specific category in the survey. Investigation and compliance are treated as workflow stages. Moderation and abuse are comparison areas, although some of the most explicit latency/cost and governance/fairness evidence in the survey comes from moderation rather than fraud (see §6). Robustness, explanation integrity, and selective prediction cut across all of these areas. The frame locates a paper's role and exposes missing evidence; it should not be read as a maturity ranking.

\subsection{Anchor Domain: Fraud Detection, Scam, and Phishing}

Fraud detection is the largest task-specific area in this corpus, but the evidence is uneven. The 18 fraud and investigation sources span direct classifiers, retrieval support, graph enrichment, explanation, and multi-step investigation. Most place LLMs as selective or supporting components around existing fraud workflows.

\paragraph{LLM as classifier.} The most direct application uses LLMs as classifiers. RL-trained LLMs for credit-card fraud \cite{RLLLMFraud}, LLM-enhanced self-evolving RL for e-commerce payment risk \cite{LLMSelfEvolvingRL}, zero-shot text-graph alignment \cite{FRAUDLLM}, telecom multi-role prompting \cite{TelecomMRML}, and telecom neuron selection \cite{TelecomNeuron} all put the LLM close to the decision boundary. These papers provide task-performance evidence under sparse-label or reasoning-heavy settings but little pipeline evidence. Latency, cost, calibration, and escalation impact are usually secondary. The published results therefore support considering these models first for offline scoring or selective second-stage use. Putting them on a real-time scoring path would require additional evidence on latency budgets, per-decision cost, calibration, and escalation impact; §6 makes those requirements explicit.

\paragraph{Retrieval-augmented decision support.} RAG gives fraud papers a clearer path to reviewable decisions. FinFRE-RAG \cite{FinFRERAG}, RAG-based phone fraud systems \cite{RAGPhoneFraud}, and evidence-grounded authentication pipelines \cite{LLMAuthFraud} treat the LLM as a component that reasons over retrieved evidence rather than as a standalone engine. This design can expose sources for inspection and support reviewer handoff. It also creates an integrity problem: retrieval poisoning \cite{PoisonedRAG} turns the evidence channel into an attack surface. RAG-based fraud systems consequently need retrieval-integrity tests alongside answer-quality metrics.

\paragraph{LLM as feature extractor and graph enhancer.} FLAG \cite{FLAG} integrates LLMs with GNNs for fraud detection, using semantic similarity neighbor sampling and LLM-based node enhancement on an Alipay-scale graph. DGP \cite{DGP} uses graph-enhanced prompting to preserve fine-grained target-node information while summarizing neighborhoods. MLED \cite{CanLLMsFraudsters} enriches graph fraud detection at type and relation levels. Here the LLM does not have to make the final decision. It can enrich representations offline, while the graph model handles the scoring path. The open evidence question is narrower and practical: do those enriched representations stay useful under drift and adversarial adaptation?

\paragraph{Scam and phishing detection.} EXPLICATE \cite{EXPLICATE} combines attribution methods with LLM-generated narratives for phishing analysis. The LLM is mainly an explanation generator rather than the primary classifier, so label accuracy does not test whether its rationale directs reviewers to the right evidence. Because scam and phishing records are often text-heavy, rationale fidelity and adversarially crafted language are immediate evaluation concerns in this part of the corpus.

\subsection{Investigation and Compliance as Workflow Stages}

LLMs are being applied to the investigative and compliance workflows that follow initial fraud signals. Compared with real-time transaction scoring, these stages often allow more processing time, but they require reasoning over heterogeneous evidence and may involve several steps.

Co-Investigator AI \cite{CoInvestigatorAI} applies multi-step LLM agents to AML compliance, generating investigative narratives that synthesize evidence across data sources. CASE \cite{CASE} applies agentic AI to scam-intelligence collection and strategic campaign analysis. The FAA Framework \cite{FAAFramework} automates credit-card fraud investigations using multi-modal LLM agents in a multi-step workflow reported as seven steps on average; in the coded evidence, its evaluation centers on evidence recovery, logical consistency, and token usage across 500 investigations rather than on latency, per-decision dollar cost, calibration, or adversarial robustness. An exploratory AML graph study \cite{AMLGraphICL} uses few-shot prompts over serialized synthetic subgraphs to classify suspiciousness and generate justifications. It describes the outputs as analyst-style logic but does not report a comparison with human analysts.

These systems produce narratives, summaries, or structured case artifacts for later review. Their evaluations need to cover completeness, evidence fidelity, analyst utility, and automation-bias risk in addition to task accuracy. Because legitimate case records can contain attacker-controlled text, indirect prompt injection also belongs in the threat model. Human-centered and adversarial tests would provide evidence that the current offline evaluations largely omit.

\subsection{Comparison Area: Moderation, Abuse, and Account Security}

Content moderation and abuse prevention provide several of the corpus's clearest operational examples. A PRISMA-following survey \cite{LLMModSurvey} covers LLM moderation from architectures to deployment practices, and the legitimacy-based evaluation framework \cite{ModAccuracyLegitimacy} argues for assessment that distinguishes easy from hard cases and extends beyond accuracy.

On the systems side, Aegis \cite{Aegis} and FlexGuard \cite{FlexGuard} use context-sensitive risk scoring or strictness controls. On 150K Reddit comments, SLM-Mod \cite{SLMMod} reports average gains of 11.5\% in accuracy and 25.7\% in recall for fine-tuned small language models ($<$15B parameters) over zero-shot LLMs. The paper discusses cost and scalability, although it does not report per-decision latency. Policy-as-Prompt \cite{PolicyAsPrompt} argues that encoding moderation policies as natural-language prompts can reduce the extensive data curation used in conventional policy operationalization. Work on LLM-enhanced personal moderation \cite{IntegratingModLLM} discusses integration with platform infrastructure.

Aegis and FlexGuard vary moderation strictness using contextual risk signals. Policy-as-Prompt moves policy changes into prompt text, whereas SLM-Mod documents a precision-recall tradeoff between smaller fine-tuned models and larger zero-shot models. SLM-Mod also tests both subreddit-specific and cross-community moderation. Although moderation is the smaller task-specific category, these papers report thresholding, policy encoding, community adaptation, and model-efficiency considerations more explicitly than most of the fraud papers.

\paragraph{Fairness and legitimacy audits.} Three fairness-focused papers in this group are from 2026. An audit across multiple LLMs reports that identical sentences rendered in African American English versus Standard American English are rated less professional and more toxic \cite{TornbergStereotypes}. Related work shows that this disparity can travel through implicit dialect signals even when explicit demographic profiles are withheld \cite{HaqDialect}. FairToT \cite{FairToT} reduces group disparities by identifying cases where demographic-related variation is likely and selectively invoking additional toxicity assessment. A separate 2025 study, AI Watchman \cite{AIWatchman}, tracks moderation refusals across model families and detects unannounced policy shifts, supporting audits of policy stability over time.

\subsection{Cross-Cutting Axes: Robustness, Explanation Integrity, and Selective Prediction}

Several concerns cut across all application areas and affect whether an LLM-based trust-and-safety system can be deployed responsibly.

\paragraph{Robustness.} Indirect prompt injection is especially relevant to trust-and-safety systems because they process untrusted content alongside system instructions or retrieved evidence. A large public competition \cite{IPIComp} reports that all 13 frontier models tested remain vulnerable at per-attempt success rates of 0.5--8.5\%, with attacks transferable across providers. Earlier survey work catalogs the attack space \cite{PromptInjSurvey}. Agent Security Bench \cite{AgentSecBench} benchmarks agent-based systems under adversarial conditions, and two agent-security surveys \cite{AgentSecInfoFusion, AgentSecACMCS} map the broader threat model. Retrieval poisoning \cite{PoisonedRAG} is the parallel risk for RAG-based fraud and investigation systems. On the Security Stack Exchange corpus, Semantic Chameleon \cite{SemanticChameleon} reports that hybrid BM25-plus-vector retrieval reduces the success of its gradient-guided poisoning attack, showing that retriever design can affect exposure to this attack.

Recent incident reports illustrate the mechanism without serving as novelty evidence. Reported Slack AI and ServiceNow Now Assist disclosures describe agent-of-the-victim failures, where an assistant acting with the user's or service's authority becomes the path for data exposure \cite{SlackExfil, ServiceNowVuln}. Microsoft's enterprise prompt-abuse guidance and OWASP's 2026 round-up describe the same broader class of indirect-prompt-injection and cloud-agent identity risks \cite{MSPromptPlaybook, OWASPRoundup}. A separate industry report describes frontier-model use in an intrusion campaign \cite{BloombergMexico}. These examples sit in the deployment-context tier of evidence: they motivate adversarial testing for T\&S workflows, but they do not by themselves establish prevalence or field-wide generalization.

\paragraph{Explanation integrity.} SHAPLLM \cite{SHAPLLM} grounds natural-language explanations in feature attributions, but work on explanation divergence \cite{LLMExplDiv} shows that LLM self-explanations can differ sharply from post-hoc attributions. CausalCoT \cite{CausalCoT} reports cases where model predictions fail to track edits to intermediate structures, suggesting that free-form explanations should not be treated as faithful without additional tests. In trust-and-safety settings, reviewers and auditors act directly on explanations, making fidelity a deployment requirement. Hallucination mitigation \cite{HallucinationMitigation} further underscores that explanation quality cannot be treated as cosmetic.

\paragraph{Selective prediction.} Abstention in an operational T\&S system is a routing decision: handle the case automatically, with assistance, or by full human review? Selective conformal risk control \cite{SCRK} offers one principled framework. Learnable conformal abstention \cite{ConformalAbstention} integrates RL with conformal prediction to optimize thresholds dynamically and reports improved performance on its hallucination-detection benchmark. Work on unequal uncertainty \cite{UnequalUncertainty} shows that uncertainty-driven routing can create fairness distortions if not carefully evaluated.

\subsection{Taxonomy Summary}

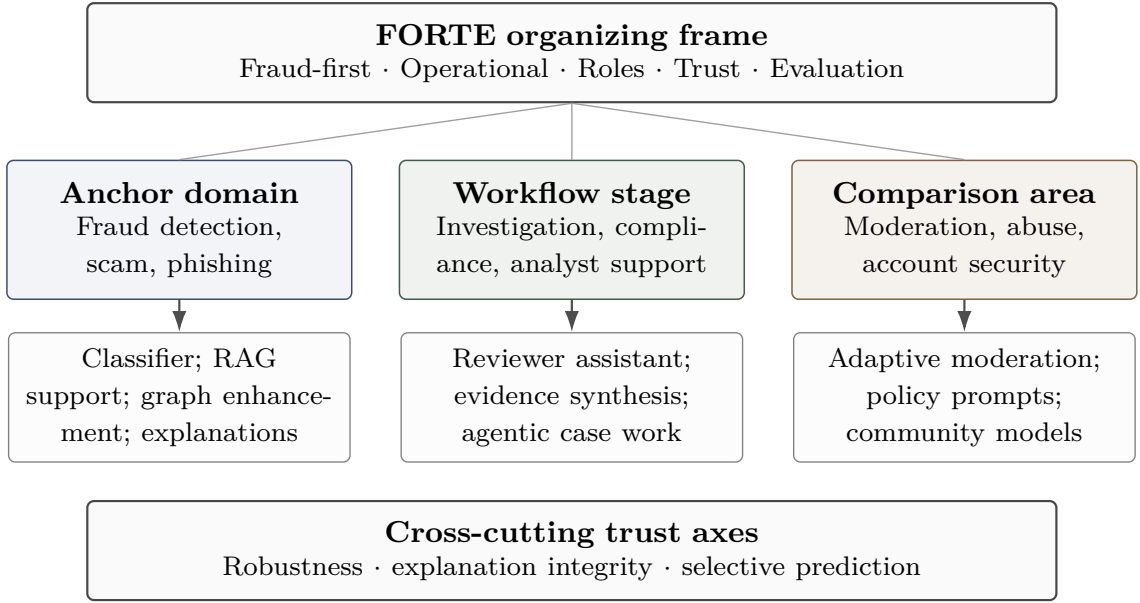
\begin{figure}[t]
\centering
\resizebox{0.95\textwidth}{!}{%
\begin{tikzpicture}[every node/.style={font=\footnotesize}]
    \node[surveybox,text width=9.7cm] (root) at (0,0) {\textbf{FORTE organizing frame}\\[-0.5mm]\scriptsize Fraud-first $\cdot$ Operational $\cdot$ Roles $\cdot$ Trust $\cdot$ Evaluation};

    \node[bluebox,text width=3.25cm,minimum height=1.45cm] (anchor) at (-4.1,-1.85) {\textbf{Anchor domain}\\[-0.5mm]\scriptsize Fraud detection, scam, phishing};
    \node[greenbox,text width=3.25cm,minimum height=1.45cm] (workflow) at (0,-1.85) {\textbf{Workflow stage}\\[-0.5mm]\scriptsize Investigation, compliance, analyst support};
    \node[orangebox,text width=3.25cm,minimum height=1.45cm] (comparison) at (4.1,-1.85) {\textbf{Comparison area}\\[-0.5mm]\scriptsize Moderation, abuse, account security};

    \node[rolebox,text width=3.25cm,minimum height=1.35cm] (anchorroles) at (-4.1,-3.6) {\scriptsize Classifier; RAG support; graph enhancement; explanations};
    \node[rolebox,text width=3.25cm,minimum height=1.35cm] (workflowroles) at (0,-3.6) {\scriptsize Reviewer assistant; evidence synthesis; agentic case work};
    \node[rolebox,text width=3.25cm,minimum height=1.35cm] (comparisonroles) at (4.1,-3.6) {\scriptsize Adaptive moderation; policy prompts; community models};

    \node[surveybox,text width=9.7cm] (cross) at (0,-5.2) {\textbf{Cross-cutting trust axes}\\[-0.5mm]\scriptsize Robustness $\cdot$ explanation integrity $\cdot$ selective prediction};

    \draw[line width=0.4pt,draw=black!38] (root.south) -- (anchor.north);
    \draw[line width=0.4pt,draw=black!38] (root.south) -- (workflow.north);
    \draw[line width=0.4pt,draw=black!38] (root.south) -- (comparison.north);
    \draw[figarrow] (anchor) -- (anchorroles);
    \draw[figarrow] (workflow) -- (workflowroles);
    \draw[figarrow] (comparison) -- (comparisonroles);
\end{tikzpicture}
}
\caption{FORTE organizing frame. Fraud detection is the anchor domain; investigation and compliance are downstream workflow stages; moderation and abuse are comparison areas. Robustness, explanation integrity, and selective prediction are cross-cutting axes.}
\label{fig:taxonomy}
\end{figure}

Table~\ref{tab:forte} summarizes the FORTE organizing frame. Figure~\ref{fig:taxonomy} provides a visual overview.

\begin{table}[ht]
\centering
\caption{FORTE Organizing Frame}
\label{tab:forte}
{\setlength{\tabcolsep}{3.5pt}
\renewcommand{\arraystretch}{1.12}
\footnotesize
\begin{tabularx}{\textwidth}{L{2.25cm}Y Y Y}
\toprule
\textbf{Lens} & \textbf{Representative papers} & \textbf{Roles and adaptation} & \textbf{Evidence / deployment reading} \\
\midrule
Fraud, scam, phishing & \cite{FLAG, FinFRERAG, RLLLMFraud, FRAUDLLM, EXPLICATE, TelecomMRML, CauPhishExp} & Classifier, RAG support, graph enhancer, explanation generator; RAG, fine-tuning, graph+LLM, prompting & Mostly offline metrics; some investigation/user-facing evidence; hybrid and escalation-aware deployment remains underreported \\
\midrule
Investigation stage & \cite{CoInvestigatorAI, CASE, FAAFramework, AMLGraphICL} & Reviewer assistant, agent, case worker; agentic workflows, ICL, retrieval & Task usefulness and evidence fidelity; assistive, human-in-the-loop deployment \\
\midrule
Moderation / abuse & \cite{Aegis, FlexGuard, SLMMod, PolicyAsPrompt, LivestreamDistill, StreamMonitor, FairToT, AIWatchman} & Classifier, policy assistant, adaptive moderator, fairness corrector; prompting, fine-tuning, thresholding, distillation & Precision/recall, legitimacy, fairness, latency/cost evidence; assistive to semi-automated production settings \\
\midrule
Cross-cutting & \cite{AgentSecBench, PoisonedRAG, SHAPLLM, LLMExplDiv, SCRK, IPIComp, SemanticChameleon, CausalCoT, PolyGuard} & Robustness benchmark, explanation verifier, abstention, RAG defense, multi-domain guardrail; adversarial evaluation, attribution, conformal prediction & Applies across all roles; evaluates robustness, explanation integrity, abstention, and transfer \\
\bottomrule
\end{tabularx}
}
\end{table}

\section{Deployment Patterns}

Knowing what an LLM-T\&S system is does not yet tell you where it sits inside a working pipeline. The evidence matrix records dominant roles for 32 application-focused sources: 18 fraud and investigation papers and 14 moderation and abuse papers. Nineteen of those sources fall into four recurring placement patterns: 10 bounded-automation papers, 3 retrieval-grounded analyst-support papers, 3 hybrid-escalation papers, and 3 multi-step investigation-support papers. The other 13 are coded as offline enrichment, reviewer or explanation support, fairness audits, deployment integration, monitoring, surveys, or evaluation frameworks.

The AML graph-reasoning pipeline \cite{AMLGraphICL} illustrates one boundary in these counts. It is coded as investigation support but not as one of the three multi-step investigation systems because it studies single-pass in-context reasoning over serialized subgraphs rather than an agentic workflow. Across the four patterns, the main difference is how much decision authority reaches the LLM before human review.

\paragraph{Pattern 1: Bounded Automation (10).} The LLM acts as a direct classifier or a tightly constrained scoring component. This is the dominant role in classifier-style work \cite{RLLLMFraud, FRAUDLLM, TelecomMRML} and is most plausible in offline scoring, selective second-stage use, or as a precursor to compression. SLM-Mod \cite{SLMMod} provides a moderation example: fine-tuned models under 15B parameters outperform zero-shot LLMs on the reported tasks. The paper discusses cost and scalability but does not report per-decision latency, and it studies fine-tuning rather than teacher-student distillation.

\paragraph{Pattern 2: Retrieval-Grounded Support (3).} The LLM is inserted after evidence retrieval to synthesize records, explain recommendations, or help a reviewer interpret heterogeneous context. FinFRE-RAG \cite{FinFRERAG}, RAG-based phone fraud \cite{RAGPhoneFraud}, and evidence-grounded authentication support \cite{LLMAuthFraud} exemplify this pattern. The AML graph-reasoning pipeline \cite{AMLGraphICL} is related but coded as investigation support rather than as one of the three dominant-role retrieval papers. In these studies, the LLM supports evidence synthesis rather than serving as the only decision component.

\paragraph{Pattern 3: Hybrid Escalation (3).} A faster component handles routine cases and invokes an LLM or human reviewer for a selected subset. In domain-knowledge-enhanced fraud detection \cite{DomainKnowledgeFraud}, a one-class drift detector flags semantic shifts and sends those conversations to a second LLM for benign-versus-adversarial classification. Bachar et al. \cite{EscalationPredictor} train a predictor on log probabilities, entropy, and attribution scores to identify moderation cases that should be sent to a human reviewer. FairToT \cite{FairToT} selectively invokes additional toxicity assessment when demographic-related variation is likely. FLAG \cite{FLAG} uses a related offline/online decomposition in which LLMs enrich features offline and graph models score downstream; the matrix treats it as offline enrichment rather than dominant-role hybrid escalation.

\paragraph{Pattern 4: Multi-Step Investigation (3).} Co-Investigator AI \cite{CoInvestigatorAI}, CASE \cite{CASE}, and the FAA Framework \cite{FAAFramework} exemplify workflows where the LLM behaves as a case worker, gathering evidence and producing a structured case artifact for review. These systems sit later in the pipeline, where processing budgets may be looser than in transaction scoring but evidence fidelity matters more. The 17 cross-cutting robustness and trustworthiness papers inform the risks and evaluation criteria for these patterns but are not included in the placement counts.

Across all four patterns, deployment involves monitoring requirements: explanation-quality checks, retrieval-corpus integrity, reviewer override analysis, adversarial probing, and performance under distribution shift.

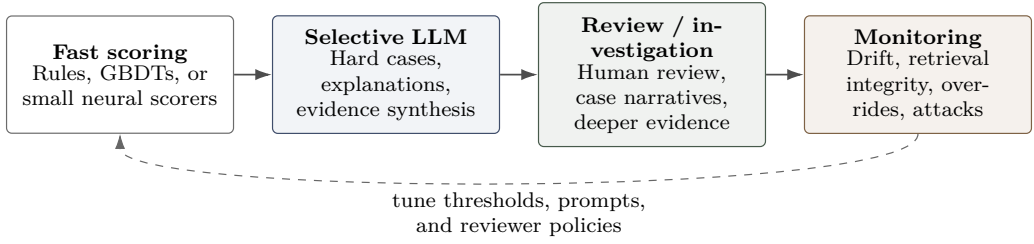
\begin{figure}[t]
\centering
\begin{tikzpicture}[node distance=6mm and 5mm, every node/.style={font=\scriptsize}]
    \node[paperbox,text width=2.65cm,minimum height=1.55cm] (fast) {\textbf{Fast scoring}\\[-0.5mm]\scriptsize Rules, GBDTs, or small neural scorers};
    \node[bluebox,text width=2.65cm,minimum height=1.55cm,right=of fast] (llm) {\textbf{Selective LLM}\\[-0.5mm]\scriptsize Hard cases, explanations, evidence synthesis};
    \node[greenbox,text width=2.65cm,minimum height=1.55cm,right=of llm] (review) {\textbf{Review / investigation}\\[-0.5mm]\scriptsize Human review, case narratives, deeper evidence};
    \node[orangebox,text width=2.65cm,minimum height=1.55cm,right=of review] (monitor) {\textbf{Monitoring}\\[-0.5mm]\scriptsize Drift, retrieval integrity, overrides, attacks};

    \draw[figarrow] (fast) -- (llm);
    \draw[figarrow] (llm) -- (review);
    \draw[figarrow] (review) -- (monitor);
    \draw[figdash] (monitor.south) .. controls +(0,-0.85) and +(0,-0.85) .. node[below,align=center,font=\scriptsize] {tune thresholds, prompts,\\and reviewer policies} (fast.south);
\end{tikzpicture}
\caption{Conceptual deployment stack for a selective architecture. Fast first-pass scoring handles routine decisions; LLMs are reserved for hard cases, explanation, and investigation. Monitoring tracks explanation quality, retrieval integrity, and adversarial behavior.}
\label{fig:deployment}
\end{figure}

\section{Evaluation Practices and Trustworthiness}

The corpus emphasizes offline benchmark metrics, while deployment decisions also require latency, cost, routing, reviewer-impact, and failure-mode evidence. Because the missing evidence differs by role, this section examines each evaluation layer separately.

The evaluation-layer coding is intentionally strict. Clean latency means per-decision latency, streaming latency, or an explicit latency budget; batch runtime and offline processing windows are coded separately. Dollar cost means a reported monetary cost, cost per decision, or deployment-volume economics sufficient to estimate one; token counts are coded separately as token-usage evidence. Calibration is counted only when a paper evaluates probability calibration, threshold reliability, or coverage-risk behavior rather than reporting accuracy alone.

\paragraph{Coverage of evaluation layers.} Manual coding of the 49 operationally relevant sources shows that, among the 18 fraud and investigation sources, none report clean per-decision latency. One has page-verified batch-runtime/deployment evidence. One reports token-usage evidence. None report per-decision dollar cost, and none report calibration. Three report human or user-facing evidence: FAA's evaluation of 500 investigations, the reviewer comparison in \cite{TelecomMRML}, and the phishing-warning experiment in \cite{CauPhishExp}. The 14 moderation and abuse sources include 4 papers with efficiency or deployment-economics evidence. The cross-cutting category focuses instead on robustness, explanation integrity, abstention, agent security, and related evaluation methods.

\paragraph{The fraud-anchor tension.} Fraud is the anchor domain because it is the largest task-specific category in the coded corpus. The clearest deployment-style evidence, however, comes from moderation: livestream A/B results, streaming token-budget results, longitudinal audits, and demographic-fairness audits with measured outcome shifts. Fraud provides more task-specific depth, whereas moderation provides more public deployment evidence. Future fraud papers would improve the record by reporting a latency or runtime budget, an inference-cost measure, an explicit decision threshold, and at least one human-impact metric. Reporting even three of these four would add deployment evidence that most coded fraud papers currently lack.

\paragraph{Predictive performance.} Static benchmark metrics dominate. They are useful for comparing model families but say little about whether an LLM belongs in a live workflow. Two moderation papers provide more concrete efficiency evidence. A livestream MLLM distilled into a lightweight classifier reports a 6--8\% reduction in user views of unwanted streams in a production A/B test \cite{LivestreamDistill}. In a separate benchmark, streaming-content monitoring achieves macro-F1 $\geq$ 0.95 while inspecting only the first 18\% of generated tokens on average \cite{StreamMonitor}.

\paragraph{Human-centered evaluation.} Reviewer-assistant systems, explanation generators, and investigation tools cannot be convincingly evaluated using predictive metrics alone. The FAA Framework \cite{FAAFramework} provides one of the few examples in the coded corpus of systematic investigation-level evaluation, assessing evidence impact and logical consistency across 500 investigations. A complementary behavioral study \cite{CauPhishExp} runs a controlled $N=750$ between-subjects experiment comparing LLM-generated feature-based and counterfactual phishing-warning explanations against manually crafted baselines, measuring click-through behavior and trust ratings. Human-centered evidence remains limited within the coded corpus.

\paragraph{Robustness and adversarial evaluation.} Agent Security Bench \cite{AgentSecBench}, PoisonedRAG \cite{PoisonedRAG}, and the prompt-injection literature \cite{PromptInjSurvey} show how adversarial risks can be tested directly. The 2026 frontier-model competition \cite{IPIComp} demonstrates indirect-prompt-injection vulnerability under large-scale adversarial testing. Agentic red-teaming work \cite{AgenticRedTeam} proposes shorter assessment cycles and mappings to OWASP, MITRE ATLAS, and NIST AI RMF. The coded corpus contains no end-to-end fraud-review protocol that combines multiple attack classes under a realistic threat model, so the robustness of published LLM-for-fraud results after deployment remains unknown.

\paragraph{Explanation-integrity evaluation.} The divergence between LLM self-explanations and attributions \cite{LLMExplDiv, SHAPLLM}, plus causal evidence that intermediate structures may not mediate final predictions \cite{CausalCoT}, matters in deployment because analysts, auditors, and compliance reviewers act on those explanations.

\paragraph{Routing and resource realism.} Abstention is a routing decision. Systems should be evaluated on which cases they escalate, how those decisions interact with queue capacity, and how uncertainty-driven patterns affect fairness \cite{UnequalUncertainty, ConformalAbstention}. The legitimacy framework for moderation \cite{ModAccuracyLegitimacy} adds another question: do affected users and reviewers see the decision as justified?

Table~\ref{tab:evalgap} summarizes the most common evidentiary mismatches.

\begin{table}[ht]
\centering
\caption{Evaluation Gap Analysis}
\label{tab:evalgap}
{\setlength{\tabcolsep}{4pt}
\renewcommand{\arraystretch}{1.12}
\footnotesize
\begin{tabularx}{\textwidth}{L{2.65cm}L{2.25cm}Y L{1.85cm}}
\toprule
\textbf{Claim} & \textbf{Reported} & \textbf{Evidence needed} & \textbf{Refs.} \\
\midrule
LLM improves fraud detection & Offline metrics & Per-decision latency, dollar cost, calibration, pipeline analysis & \cite{FLAG, FRAUDLLM, FinFRERAG} \\
LLM helps reviewers & Case-level task evaluation & Controlled reviewer outcomes and workflow impact & \cite{FAAFramework, CoInvestigatorAI} \\
Rationale supports decision & Generated explanations and attribution comparisons & Faithfulness and source alignment & \cite{SHAPLLM, LLMExplDiv, EXPLICATE} \\
RAG supports fraud review & Task success & Retrieval-integrity and prompt-injection evaluation & \cite{PoisonedRAG, AgentSecBench, CASE} \\
Selective invocation viable & Confidence scores & Coverage-risk curves, fairness analysis & \cite{SCRK, ConformalAbstention, UnequalUncertainty} \\
Moderation decisions are justified & Accuracy benchmarks & Legitimacy, cross-community transfer & \cite{SLMMod, ModAccuracyLegitimacy, PolicyAsPrompt} \\
\bottomrule
\end{tabularx}
}
\end{table}

\paragraph{A minimum deployment-evidence checklist.} Table~\ref{tab:evalgap} gives the diagnostic version of the evaluation gap. The five questions below cover the evidence areas emphasized by this survey; they are a minimum, not a complete governance or deployment review.

\begin{enumerate}[nosep]
\item \textbf{Latency budget.} What end-to-end latency must the workflow meet, and what fraction of cases will invoke the LLM? Patterns 1 and 3 from §5 depend on having a clear answer.
\item \textbf{Cost per decision.} At the expected volume, what does each LLM invocation cost, and does its incremental benefit justify that cost? \cite{LivestreamDistill, SLMMod} offer partial comparators.
\item \textbf{Explicit decision threshold.} What does the LLM's output trigger automatically, what is escalated, and what receives no automated action? Performance-predictor work \cite{EscalationPredictor} and conformal abstention \cite{SCRK, ConformalAbstention} provide relevant starting points.
\item \textbf{Explanation-integrity floor.} Are rationales supported by retrieved or tool-derived evidence, and can tests detect fabricated or unfaithful explanations? This is one of the weakest evidence areas in the coded corpus. The front-door interventions in CausalCoT \cite{CausalCoT} show why the question matters, but they remain a research method rather than a routine operational test.
\item \textbf{Adversarial pressure.} What does the system do under indirect prompt injection, retrieval poisoning, and the agent-of-the-victim patterns in §4.4? \cite{IPIComp, SemanticChameleon, AgenticRedTeam} are the closest published red-team artifacts.
\end{enumerate}

Calibration belongs with item 3 because a decision threshold is only as reliable as the probabilities beneath it \cite{SCRK, ConformalAbstention}. Human impact is not fully captured by these five questions and should be reported separately through reviewer accuracy, workload, overreliance, and user outcomes. Explanation integrity remains a cross-cutting trust axis rather than a separate evidence layer.

A system missing most of this evidence remains a prototype for the purposes of this survey. The checklist gives practitioners a literature-grounded starting point for judging whether an LLM-supported workflow has enough public evidence to be considered operationally mature.

\paragraph{A worked example: FAA against the checklist.} The FAA Framework \cite{FAAFramework} is one of the most detailed investigation examples in the corpus. It provides systematic evidence over 500 cases and reports token usage, investigation-step counts, evidence impact, relevance, and logical alignment. Its published evaluation does not cover latency, per-decision dollar cost, calibration, explicit deployment thresholding, or adversarial testing. FAA therefore supplies substantial case-level evidence while leaving several deployment questions unanswered.

\section{Open Challenges and Research Agenda}

The searched literature is active but fragmented. LLMs are being applied to fraud detection, investigation, scam analysis, and moderation, yet the coded corpus lacks shared evaluation standards, systematic adversarial testing, and explicit reporting of deployment constraints.

\subsection{Cross-Cutting Themes}

\paragraph{The convergence of detection and investigation.} Many conventional fraud pipelines separate real-time scoring from later case review. RAG-based systems combine part of that work by retrieving evidence and generating explanations during scoring, while Co-Investigator AI \cite{CoInvestigatorAI} and the FAA Framework \cite{FAAFramework} automate several evidence-gathering steps. Evaluations of these combined workflows must account for component failures and specify which decisions remain with human reviewers.

\paragraph{Publication-safe deployment research.} Organizations running fraud systems may withhold operational details that could help adversaries. This likely contributes to the corpus's emphasis on academic benchmarks over realistic deployment evaluations. Publication norms modeled on responsible disclosure could make it easier to report latency, cost, reviewer outcomes, and failure modes without exposing exploitable controls.

\paragraph{Open gaps.} A targeted May 2026 search identified three questions that remain largely unaddressed in the searched corpus and window. The search combined the gap-area term (\textit{concept drift}, \textit{procedural justice}, \textit{appeals}, \textit{contestability}, \textit{cross-domain transfer}) with the LLM-T\&S vocabulary from §3 across arXiv and Google Scholar from 2024-01 to 2026-05. An ancillary search note included with the source bundle records the query families and screening rules. This was a targeted gap check rather than a systematic review; the claims below mean ``not found inside that searched window and corpus,'' not ``provably absent in all venues.''

\paragraph{Concept drift.} Only the domain-knowledge-enhanced fraud study \cite{DomainKnowledgeFraud} engages drift directly in the coded corpus. The other directly relevant examples found in the targeted search were largely vendor or industry reports.

\paragraph{Procedural justice and appeals.} Demographic-fairness audits \cite{TornbergStereotypes, HaqDialect} are represented in the corpus, but within the searched corpus and window we did not find work that moves beyond the legitimacy framing of \cite{ModAccuracyLegitimacy} to study the design, evaluation, or outcomes of appeals processes for LLM-assisted moderation.

\paragraph{Cross-domain transfer.} Multi-domain guardrail benchmarks cover fraud, phishing, and moderation together \cite{PolyGuard}, but within the searched corpus and window we did not find a study measuring whether methods trained on one trust-and-safety domain generalize to another. These three gaps anchor the research agenda below.

\paragraph{Signals after the search window.} The coded corpus and its counts are frozen at the May 2026 update. A July 2026 spot check found newer arXiv papers reporting calibration and cost-aware routing \cite{UCCIRouting}, a payments-dispute guardrail with an explicit latency budget \cite{DisputeGuardrail}, and an efficient latent-reasoning guardrail \cite{CoLaGuard}. It also found a benchmark-only fraud evaluation \cite{SAGEFraud} and a contextual-integrity argument for accepting and monitoring some prompt-injection risk rather than trying to eliminate it \cite{CIPromptInj}. These five sources arrived after the coding freeze and are not counted in the corpus statistics of §3.

\subsection{Research Agenda}

Table~\ref{tab:agenda} presents nine study designs that would fill specific gaps in the coded evidence.

The study designs require different levels of access. Deployment-realism and adversarial tests can begin with public data and existing benchmarks. Reviewer-impact studies need controlled experiments, while drift monitoring and appeals research require longitudinal or organizational access. Cross-domain transfer can start with paired public datasets but still needs common tasks and metrics. Together, these studies would bring latency, cost, human impact, and governance evidence into fraud evaluations more consistently.

\begin{table}[ht]
\centering
\caption{Concrete Next Studies for Deployment Evidence}
\label{tab:agenda}
{\setlength{\tabcolsep}{4pt}
\renewcommand{\arraystretch}{1.05}
\footnotesize
\begin{tabularx}{\textwidth}{L{1.9cm}Y L{2.85cm} L{1.45cm}}
\toprule
\textbf{Gap} & \textbf{Concrete next study} & \textbf{Success measure} & \textbf{Starting points} \\
\midrule
Deployment realism & A fraud benchmark that reports accuracy, latency, cost per decision, calibration, and review-load change on the same tasks & A Pareto frontier showing when LLM invocation is worth its cost & \cite{FLAG, SLMMod, StreamMonitor} \\
Adaptive adversaries & A red-team protocol for RAG fraud review that combines indirect prompt injection, retrieval poisoning, and attacker adaptation over repeated rounds & Failure rates by attack class and recovery after mitigation & \cite{PoisonedRAG, IPIComp, AgenticRedTeam} \\
Explanation integrity & A fraud-review study comparing free-form rationales, attribution-grounded rationales, and tool-derived explanations under auditor inspection & Reviewer detection of fabricated or unsupported rationales & \cite{SHAPLLM, LLMExplDiv, CausalCoT} \\
Human review impact & A controlled reviewer study measuring whether LLM case summaries improve accuracy, time-to-decision, disagreement, and overreliance & Net reviewer benefit after accounting for automation bias & \cite{FAAFramework, CoInvestigatorAI, CauPhishExp} \\
Model specialization & A latency-fixed comparison of frontier LLMs, distilled moderation-style models, and domain-tuned fraud models & Equal-latency accuracy, cost, and escalation trade-offs & \cite{SLMMod, LivestreamDistill, RLLLMFraud} \\
Selective prediction & A routing benchmark where systems choose automatic decision, LLM review, human review, or abstention under asymmetric costs & Coverage-risk-cost curves with fairness slices & \cite{SCRK, ConformalAbstention, UnequalUncertainty} \\
Monitoring and drift & A longitudinal fraud or moderation evaluation that tracks threshold drift, reviewer overrides, retrieval-corpus changes, and adversarial probes & Drift alarms linked to action thresholds, not just dashboard plots & \cite{DomainKnowledgeFraud, AIWatchman} \\
Cross-domain transfer & A paired fraud/moderation benchmark testing whether prompts, guardrails, abstention rules, or explanation methods transfer across domains & Transfer loss relative to in-domain tuning & \cite{PolyGuard, LLMModSurvey, PolicyAsPrompt} \\
Appeals and contestability & A human-centered study of LLM-assisted moderation or fraud appeals, including explanation quality, perceived legitimacy, and reversal outcomes & Appeal accuracy and perceived procedural fairness, not only first-pass accuracy & \cite{ModAccuracyLegitimacy, TornbergStereotypes, HaqDialect} \\
\bottomrule
\end{tabularx}
}
\end{table}
\clearpage

\paragraph{Using the survey.} Researchers can use FORTE to position new methods by deployment role. Practitioners can use it to identify which LLM roles have public support and which lack the evidence to back deployment claims. Benchmark and robustness authors can use the evaluation-gap analysis (Table~\ref{tab:evalgap}) as a checklist of evidence layers that should appear together in future high-stakes LLM studies.

\section{Conclusion}

Fraud detection and adjacent trust-and-safety uses of LLMs are systems problems. Treating them that way changes which evaluations matter.

The role-and-evidence frame organizes the field by deployment role, not by model family. In the surveyed work, LLMs appear as classifier-like decision support, evidence synthesizers, reviewer assistants, explanation generators, graph enhancers, and multi-step investigation tools. The corpus reports offline metrics far more consistently than deployment realism. Abstention, adversarial robustness, explanation fidelity, and human-AI interaction are still under-reported. Cost and latency are starting to be reported in moderation. Among the fraud and investigation sources, one has page-verified batch-runtime/deployment evidence and FAA reports token usage, but clean per-decision latency, per-decision dollar cost, and calibration do not appear in the coded evidence.

The evidence supports selective and hybrid architectures as the most defensible near-term option. In these designs, faster conventional models handle routine decisions and LLMs are reserved for explanation, escalation, or investigation. The corpus does not establish whether an LLM can replace an existing real-time fraud engine. Evidence for such a claim would need to cover latency, cost, calibration, and escalation impact.

The research agenda prioritizes evidence that changes deployment decisions. Latency, cost, calibration, and adversarial tests can strengthen system evaluations immediately. Reviewer-impact and selective-routing studies require more infrastructure but would show how model behavior affects actual work. Cross-domain transfer and appeals remain the least developed areas in the coded corpus and the 2024-01 to 2026-05 targeted gap search.

\bibliographystyle{unsrtnat}
\bibliography{references}

\end{document}